\documentclass{PoS}

\def \aap{A\&A }
\def \apjl{ApJ}

\def \apj{ApJ}

\def \mnras{MNRAS}

\def \nat{Nature\ }
\def\msun{{\,M_\odot}}

\title{Non-Thermal Radiation from the Inner Galaxy}

\ShortTitle{Non-Thermal Radiation from the Inner Galaxy}

\author{\speaker{Roland Crocker}\thanks{ARC Future Fellow}\\
        Research School of Astronomy and Astrophysics, Australian National University\\
        E-mail: \email{rcrocker@fastmail.fm}}

\abstract{I review our current state of knowledge about non-thermal radiation from the Galactic Centre (GC) and Inner Galaxy.
Definitionally, the Galactic nucleus is at the bottom of the Galaxy's gravitational well, rendering it a promising region to seek the signatures of dark matter decay or annihilation.
It also hosts, however, the Milky Way's resident supermassive black hole and up to 10\% of current massive star formation in the Galaxy.
Thus the Galactic nucleus is a dynamic and highly-energized environment 
implying that extreme caution must be exercised in interpreting any unusual or unexpected signal from (or emerging from) the region as evidence for dark matter-related processes.
One spectacular example of an `unexpected' signal is the discovery within the last few years of the `Fermi Bubbles' and, subsequently, their polarised radio counterparts.
These giant lobes extend $\sim$7 kpc from the nucleus into both north and south Galactic hemispheres.
Hard-spectrum, microwave emission coincident with the lower reaches of the Bubbles has also been detected, first in WMAP, and more recently in Planck data.
Debate continues as to the origin of the Bubbles and their multi-wavelength emissions: are they the signatures of relatively recent (in the last $\sim$Myr) activity of the supermassive black hole or, alternatively, nuclear star formation?
I will briefly review evidence that points to the latter interpretation.
}

\FullConference{Cosmic Rays and the InterStellar Medium - CRISM 2014,\\
		24-27 June 2014\\
		Montpellier, France}

\begin{document}

\section{Introduction and Cautionary Remarks}

In this brief review I will use `Galactic Centre' (GC) to mean the region of $\sim$300 pc diameter across the Galactic plane surrounding the Galaxy's super-massive black hole (SMBH) at Sgr A$^*$.
As the bottom of the Galaxy's gravitational well, the GC should be the region displaying the brightest radiative signature of dark matter decay/annihilation in the Galaxy \cite{Bergstrom1998}.
Yet, because it enjoys this privileged geometrical position, the GC always accretes gas at some level: any process that generates negative toque on gas 
further out in the Galactic disk will tend to send this gas inwards where it might, in principle, either be accreted onto the SMBH or, as seems to be happening in today's GC, 
fuel the formation of new generations of stars. 
In fact, the evidence seems to be that the GC is currently forming stars at a rate approaching 0.1 $\msun$/year \cite{Crocker2012}, consistent with the rate of nuclear star-formation averaged over the life of the Galaxy \cite{Serabyn1996,Figer2004}.

Galactic stellar surface density and current star-formation actually reach a strong crescendo in the GC.
This means that ISM conditions in the region are unusual in terms of the rest of the Galaxy: the various phases of the GC ISM 
(dense and highly turbulent molecular gas; diffuse, hot plasma; magnetic fields; cosmic rays) have energy densities  $\sim$2 orders of magnitude larger than typical for the Galaxy-at-large\cite{Crocker2012}.
One must then contemplate that there will be phenomena that occur  in the GC (or in the Inner Galaxy, driven by processes in the GC) that, while unique to that region of the Galaxy, can be explained ultimately as due to the high star-formation rate densities or the SMBH.
Any would-be dark matter decay or annihilation signal, then, occurs with such unusual but `conventional' astrophysical processes as a co-located background.
Thus, extreme caution must be exercised in adducing an unusual GC signal as evidence, say, for dark matter initiated processes.
To put this another way, to claim detection of a dark matter signal, we have to subtract off the astrophysical backgrounds. 
Given the unusual nature of the GC environment, we should be cautious about training our expectations for such backgrounds by reference to observations elsewhere in the Galaxy.

\section{The GC Environment and its Denizens}

From our perspective, the GC is viewed though the Galactic plane and along a line of sight quite close to the long axis of the Galactic bar.
This  results in many order of magnitude in visual extinction with the implication that, if we want to observe the nucleus directly we have to go to wavelengths either significantly longward or shortward of the optical band.
Much of our information about the GC, then, relates to non-thermal processes which are often the dominant sources of radiation at $\sim$ GHz radio frequencies, on the one hand, and in high-energy X-rays and $\gamma$-rays on the other.

Observations of molecular line emission from the GC  in radio bands and of dust emission at infrared wavelengths (e.g., \cite{Morris1996,Molinari2011}) inform us that the GC contains about $3 \times 10^7 \ \msun$ of molecular hydrogen, about 10\% of all such star-forming fuel in the Galaxy.
Much of this gas is associated to a twisted, somewhat irregular torus that circles (the somewhat offset) supermassive black hole at a radius of $\sim 100$ pc \cite{Molinari2011}.
This structure may be an analogue of the (often much larger) nuclear star-forming rings found in external, (face on) barred, spiral galaxies.
Given the GC's allocation of $H_2$, it is responsible for a reasonable fraction of the massive star formation in the Galaxy, $5-10$ \% \cite{Morris1996}.
%
%
The evidence seems to be that the GC is populated with stars with a tremendous range of ages: 
from massive and short-lived Wolf Rayet stars and Luminous Blue Variables to red giants requiring progenitors of many Gyr age.

The GC also contains some of the densest and most massive clusters in the local universe: the nuclear stellar cluster surrounding the the SMBH and the Arches and Quintuplet clusters
that orbit at a distance of $r < 200$ pc (and are both currently only $\sim 30$ pc in projection from the SMBH).
These clusters have total masses
$\sim 10^4 \msun$, probably rather heavier at birth, and contain many O and B stars (see \cite{Stolte2014} and references therein).
Both Arches and Quintuplet, with ages of a few Myr, have likely already or will soon experience core collapse supernova explosions as their most massive members expire; 
indeed there is some gas dynamics and radio continuum evidence for such explosions occurring in the vicinity of Quintuplet \cite{Sofue2003}.
The strong tidal forces in the GC gradually disperse the orbiting clusters \cite{Stolte2014} until they become invisible against the heavily populated background of the nuclear bulge stars \cite{Serabyn1996}
so the GC has likely experienced the creation of many previous generations of individual clusters in mini-starbursts over its lifetime.
The exact mechanics behind the creation of these orbiting clusters remains under debate \cite{Longmore2013}.

\section{Exotic/Remarkable Non-Thermal Phenomena of the GC/Inner Galaxy}

The GC exhibits many remarkable or unusual phenomena.
Some of these were recently discovered, but many have been known of, but still not understood, after many decades.
These phenomena include:
\begin{itemize}

\item The (quasi) point-like GeV and TeV $\gamma$-ray source coincident with Sgr A* (= radio source coincident with the SMBH) \cite{Aharonian2004,Aharonian2009,Chernyakova2011}.

\item
GeV and TeV  \cite{Aharonian2006} emission extended (few degrees) along the Galactic plane (the `ridge')

\item
A few GeV $\gamma$-ray spectral bump (\cite{Daylan2014} and references therein) distributed in an apparently spherically-symmetric fashion around the GC and extending to $\sim 10^\circ$ radius

\item
Non-Thermal Radio (and X-ray) Filaments (NTFs) \cite{Yusef-Zadeh1984}

\item
130 GeV `line' \cite{Weniger2012}

\item
The 511 keV positron annihilation line coincident with the Bulge \cite{Prantzos2011}

\item
The non-thermal microwave `haze' \cite{Finkbeiner2004}

\item
The Fermi Bubbles \cite{Dobler2010,Su2010}

\end{itemize}
It is remarkable in itself that {\it every single one of these phenomena has been claimed as potential evidence of dark matter-related processes}.
I discuss some of these in more detail below

Another long-standing  anomaly (or, at least, bone-of-contention) connected to the GC is that X-ray measurement of the 6.7 keV Fe line
suggest that the inner $\sim 300$ pc  is suffused by a very hot, $\sim$ 7 keV plasma \cite{Koyama1989,Muno2004}.
Were this plasma real, it would quickly escape the nuclear region and would seem to require  heating  unsupportable with the current supernova rate.
It may be, however, that this apparently diffuse signal is really attributable to many faint, unresolved point sources, presumable white dwarfs and coronally active stars \cite{Revnivtsev2009} but there remains active debate on this question \cite{Uchiyama2013}.
To my knowledge this signal has not been interpreted as evidence for dark matter (I may be wrong) 
but a more recent anomaly connected with a mysterious X-ray line at 3.53 keV, seen towards the GC, Andromeda and come galaxy clusters, has been: see \cite{Boyarsky2014} and references therein.

\section{Energetics}

For reference, I list here some relevant numbers for the energetics associated with various GC objects/processes.
The (photon) Eddington luminosity of the SMBH ( with $4 \times 10^6 \msun$) is $5 \times 10^{44}$ erg/s.
Given star formation in the CMZ proceeds at a time-averaged rate $\sim 0.1 \msun$/year \cite{Crocker2012},
it injects mechanical power (supernova explosions, stellar winds) of 
$P_{mech} \sim 0.1 \msun$/year $ \times 1 SN/(90 \msun) \times 10^{51}$ erg/SN 
  = $4 \times 10^{40}$ erg/s.
Here I have normalized to 1 core-collapse supernova per 90 $ \msun$ of formed stars (for a Kroupa IMF with a minimum 8 $\msun$ zero age main sequence stellar mass required for a star to end its life 
as in a core collapse explosion: see \cite{Crocker2012} and references therein) and assumed $10^{51}$ erg per SN.

\section{The Central $\gamma$-ray Source}

The central source has a complicated spectrum and is not describable as a pure power law, a fact that becomes especially evident if one 
naively connects the Fermi and HESS spectra.
In particular apparent bumps in the spectrum can be interpreted as being characteristic of the annihilation of dark matter particles with mass scales of 10-1000's of GeV (e.g., \cite{Belikov2012}).
The dark matter explanation is not unique, however, and `conventional' explanations are possible (and plausible) for the overall spectrum.
For instance, the steepening between the Fermi and HESS spectral regions may be due to a transition in the nature of cosmic ray proton transport in the vicinity of the SMBH \cite{Chernyakova2011}.
Imagine, in particular, that at  low energies such particles are effectively trapped in this region and lose all their energy in situ to hadronic collisions. 
Then, in this `thick target' regime, the $\gamma$-ray spectrum from the region simply reflects the hard injection distribution of the protons.
At somewhat higher energies, the particles start to escape diffusively on a timescale less than the $pp$ loss time and -- as for  Galactic plane CRs -- 
this leads to a steepening of the in-situ cosmic ray distribution because of the growth of the diffusion coefficient with energy.
The  steepening of the proton spectrum is mirrored by the $\gamma$-ray secondaries produced by the diffusing cosmic rays.
Finally, at the highest energies (leading up to an exponential cut-off presumably reflecting the maximum energy attainable within the accelerator) particles escape rectilinearly at the speed of light (and no faster)
so that, once again, transport is energy {\it independent} resulting in a hard (but relatively attenuated) $\gamma$-ray spectrum determined by the injection distribution solely.

\section{The  $\gamma$-ray Ridge}

On larger scales, up to $\sim 2^\circ$, Fermi and HESS observations point to diffuse $\gamma$-ray emission extended along the Galactic plane and correlated with the column of molecular gas \cite{Aharonian2006,Yusef-Zadeh2013,Macias2014}.
At $\sim$TeV energies the emission from this `Ridge' is hard, $ \propto E_\gamma^{-2.3}$ or so, similar to the spectrum of the TeV point source of top of Sgr A$^*$; at GeV energies the spectrum may be softer, $ \propto E_\gamma^{-2.7}$ more-or-less consistent with expectation were emission to be created by
bombardment of the region's molecular gas by the same Galactic `sea' of cosmic rays that permeates the disk on wider scales \cite{Yang2014}.
In general, because of the  spatial correlation between the observed diffuse $\gamma$-rays and the molecular gas column, it has generally been supposed that this emission arises from hadronic cosmic rays (protons and heavy ions) though an alternative is that the emission is actually bremmstrahlung emission from cosmic ray electrons
\cite{Yusef-Zadeh2013} though such requires that the magnetic field suffusing the GC be surprisingly low.
Interestingly, there is  extended, diffuse $\sim$GHz radio emission also roughly correlated with the $\gamma$-rays on the same spatial scales \cite{LaRosa2005,Crocker2010,Yusef-Zadeh2013}.
A separate question is whether this emission represents a steady state or, alternatively, some recent outburst presumably from or associated with Sgr A$^*$ or supernovae in its vicinity\cite{Aharonian2006,Macias2014b}.
 
\section{The  Few GeV Bump}

In addition to the ridge, the detection of a startling spectral/morphological feature in the few GeV band around the GC has been claimed in a number of papers over the last few years, led by Dan Hooper and co-workers 
(see \cite{Hooper2011} and
\cite{Daylan2014} and references therein).
This is a spectral bump peaking at around $\sim 2$ GeV that is centrally concentrated around the GC.
Tantalizingly, the bump seems to die off in intensity in a fashion consistent with expectation were it due to dark matter annihilation where the dark matter profile is governed by law $\rho_{DM} \propto r^{-\gamma}$ where $1.1 < \gamma < 1.3$ and with a `reasonable' velocity-averaged cross-section $\sigma v = (1.4-2.0) \times 10^{-26}$ cm$^3$/s (normalized to a local dark matter density of 0.3 GeV/cm$^3$).
This signal, which has been detected out to radial scales of $\sim 10^\circ$, is indeed intriguing.
One potential astrophysical origin for it is a population of previously-unknown and currently unresolved millisecond pulsars in the region \cite{Hooper2011,Abazajian2011,Gordon2013}.
The spectra of individual millisecond pulsars and, indeed, of globular clusters (whose $\sim$GeV emission is thought to be dominated by MSPs) does resemble the GC bump feature.
There have been claims and counter-claims regarding whether the MSP idea stacks up in detail, however (see \cite{Daylan2014} and references therein).
Note that a DM candidate distributed like $\propto r^{-1.3}$ would release energy in annihilation products at a rate of $3 \times 10^{37}$ erg/s within the innermost 150 parsecs around the Galactic Centre.
Within the same region, however, star-formation (supernovae, stellar winds) injects $\sim10^{39}$ erg/s in accelerated hadrons and $\sim10^{38}$ erg/s  in accelerated leptons (i.e., cosmic ray electrons) - energetically this is thus a reasonable match to the  bump signature (though does not necessarily explain its shape).
Preliminary attempts to explain the $\gamma$-ray bump with diffuse cosmic ray populations have started to appear: see \cite{Carlson2014,Petrovic2014}.

\section{The  Positron Signal}

In contrast to the relatively-recently discovered bump feature, another mysterious, non-thermal feature from the inner Galaxy has been known for forty years (and has been causing theorists to scratch their heads for a similar length of time).
This is the 511 keV line and continuum emission detected in soft X-rays that betokens the annihilation of around $\sim 10^{43}$ positrons per second in the Galactic bulge \cite{Prantzos2011}.
The origin of these positrons is still unknown.
Interestingly, the positron annihilation signal is similar in extent to the distribution of the bump, perhaps suggesting some connection between the two \cite{Boehm2014}.
$\gamma$-ray constraints imply that the positrons have to be injected into the ISM at $\sim$ few MeV energies \cite{Beacom2006} which significantly constrains many models including various dark matter-related ideas.
This energy scale is consistent with injection of positrons from $\beta^+$ decay of unstable radionuclei, but there is no obvious, conventional source of such nuclei that shares the observed distribution of the positron annihilation radiation.
For the moment, the positron mystery endures.

\section{Existence of a Nuclear Outflow}

Careful spectral analysis  \cite{Churazov2011} suggests that the positrons are born into a hot ($\sim 10^6$ K), volume-filling medium but do not annihilate until this has cooled (either radiatively or via adiabatic expansion) to $\sim 10^5$ K.
Although this does not explain the origin of the positrons, it is a point of connection to significant other evidence that strongly indicates that the Milky Way's nucleus launches an outflow into both north and south Galactic hemispheres \cite{Sofue2000,Bland-Hawthorn2003}.

Interestingly, non-thermal data strongly support the existence of an outflow.
On the basis of the star-formation activity occurring in the GC, we have an expectation for how bright the region should be
in the $\sim$ GHz radio band (where emission is dominated by synchrotron emission from cosmic ray electrons) and in the $\sim$TeV $\gamma$-ray band (where emission is dominated by hadronic emission from cosmic ray protons and heavier ions; i.e., collisions between such particles and ambient gas leading to the production of neutral mesons that decay immediately into $\gamma$-rays). 
While extended and diffuse radio and $\gamma$-ray emission has been detected around the GC, in both wavelength bands {\it the GC is much dimmer than expectation} \cite{Crocker2011a,Crocker2011b}.

This deficit is neatly explained if the large-scale outflow carries away the cosmic rays (both electrons and protons) accelerated in the region before they can radiate in situ.
This is an interesting point of difference with respect to cosmic ray transport processes elsewhere in the disk.
From the latter cosmic rays escape via diffusion by scattering on magnetic field inhomogeneities (governed by some characteristic Kolmogorov-like distribution) of a size similar to their energy-dependent gyro radii.
This means that higher energy particles escape more quickly and, thus, the steady state in situ cosmic ray distribution in the Galactic disk is steepened with respect to the injection distribution emerging from astrophysical accelerators.
In contrast, in the case of the GC, the diffuse $\sim$GHz radio continuum and $\sim$TeV $\gamma$-ray 
evidence is that in situ non-thermal particle population are not so steepened, consistent with the action of an energy-independent transport process \cite{Crocker2011a,Crocker2011b}.
A natural question is then to ask: {\it what happens to these cosmic rays next?}

\section{The Fermi Bubbles}

Perhaps the most startling discovery in high-energy astrophysics over the last decade 
and one of the highlights of the $Fermi$ mission has been the discovery \cite{Dobler2010,Su2010,Ackermann2014} of the $\gamma$-ray Bubbles that hang above and below the Milky Way nucleus, extending up to heights of $\sim 7$ kpc (i.e., approaching the distance that separates the Solar System from the GC).
These structures, which were detected by independent researchers in the Fermi data on the Inner Galaxy, 
are suffused by hard-spectrum radiation that is rather uniform in intensity, is spectrally constant (or may even harden towards high-latitudes \cite{Yang2013}), 
and apparently bounded by a sharp edge.
Counterparts to the Bubbles -- or some element of their substructure -- have been found at other wavelengths.
Hard-spectrum ($F_\nu \propto \nu^{-0.55}$), {\it total-intensity} microwave emission has been found in the lower reaches of the Bubbles (reaching to $b \sim |35^\circ|$), first in WMAP \cite{Finkbeiner2004,Dobler2008,Dobler2012} and more recently in Planck data \cite{Ade2012}.
Soft spectrum ($F_\nu \propto \nu^{-1.1}$) {\it polarised} emission from the giant, polarised radio lobe counterparts was recently discovered by the Parkes radio telescope in Australia \cite{Carretti2013} at 2.3 GHz.
The lower edges of the Bubbles appear as rather dim features in all-sky ROSAT maps; X-ray scans across the edges at high latitudes have revealed only very dim features \cite{Kataoka2013}. 

The interrelationship between all these signals is not yet clear.
More broadly, we do not know for certain where the Bubbles originate, though it is certainly some source or process occurring in the nucleus given the Bubbles' tight waist in the plane around this region.
One obvious candidate is activity in the recent past  ($< 1$ Myr ago) of the super-massive black hole: we know from observations of many external galaxies that active black holes can blow giant
radio bubbles substantially larger and more energetic than the the Fermi Bubbles.
The Fermi Bubbles probably required an energy of $\sim 10^{56}$ erg to create; this can easily be supplied by the SMBH.

But other sources of processes (too numerous to mention here) may actually have been responsible.
For instance, a scenario that I have been pursuing for some time with co-workers \cite{Crocker2011c,Crocker2012,Crocker2014} is that nuclear star-formation can inflate the Bubbles provided that the power in hot gas, cosmic rays and magnetic fields it supplies can be integrated over sufficiently long timescales.
Energetically, this idea has the attractive feature that the power we can  infer is currently being supplied by the nucleus is a good match to the luminosity (in various wavebands) of the Bubbles; thus nuclear star formation can well explain the Bubbles in a quasi steady state model.
In fact, this neatly connects to the idea presented above that the nucleus appears to be missing $\gamma$-ray and radio emission: the radio continuum and $\gamma$-ray fluxes detected from the Bubbles, are, within uncertainties, equal to the amount of flux `missing' from the GC region \cite{Crocker2012,Carretti2013,Crocker2014}.

Questions of timescales connect back to the microphysics of the $\gamma$-ray emission process: generically, the Bubbles' $\gamma$-ray emission may be largely due to inverse-Compton (IC) up scattering by a putative cosmic ray electron population of ambient light.
Alternatively, it may be largely due to  hadronic collisions between a putative cosmic ray proton and heavier ion population and the Bubbles' gas content.
Rather different timescales are associated to these processes: for IC $\gamma$-rays, dominantly off the CMB, the requisite ($\gtrsim$ TeV) electrons cool in a timescale less than 1 Myr so this mechanism naturally relates to SMBH activity because the nucleus has to supply all the Bubbles' energy content over this sort of timescale.
For the protons, the timescales are {\it much} longer: if the only gas available to collide with is the low-density Bubble plasma ($n \lesssim 10^{-2}$ cm$^{-3}$), the $pp$ loss time blows out to few $\times$ Gyr.
A steady state model thus requires that the Bubbles integrate the power fed in by nuclear star formation over the history of the Galaxy.
Recently, however, it has become clear \cite{Crocker2014} that this long timescale is avoidable: the Bubbles' thermal plasma is subject to local thermal instability leading it to collapse into local condensations and, in the process, adiabatically compressing the cosmic ray protons and magnetic fields suffusing the plasma.
The natural concentration of dense gas and amplified cosmic rays {\it together} via this process implies a revised timescale for a hadronic model of the Bubbles, more-or-less the gas cooling time of few $\times 10^8$ year.
 
 Despite these interesting theoretical ideas (and others present in the literature) a knock-down argument about the nature, age, and origin of the Bubbles is missing.
 For the moment, the Bubbles thus remain a tantalising enigma and I can only advise the reader: {\it watch this space}.

\end{document}